\documentstyle[aps,prb,epsf]{revtex}

\begin{document}
\draft
\title{Effect of the Kondo correlation on Shot Noise in a Quantum Dot}
\author{Bing Dong and X. L. Lei}
\address{Department of Physics, Shanghai Jiaotong University, 1954 Huashan Road, Shanghai 
200030, P. R. China}
\maketitle
\begin{abstract}
The current noise in a quantum dot coupled to two leads is investigated in the Kondo 
regime with and without the influence of magnetic fields by employing a finite-$U$
slave-boson mean field approach to calculate the current-current correlation function at 
zero temperature. The numerical results show that the Fano factor is always reduced 
significantly due to the Kondo-correlation effect, and the most pronounced suppression 
appears at the electron-hole symmetry case $2\epsilon_d+U=0$. In addition, the application 
of a magnetic field enhance the Fano factor at small external voltage, which is attributed 
to the reduction of the Kondo-enhanced density-of-state and transmission probability in a 
quantum dot.  
\end{abstract}

\section{Introduction}
Since the great breakthroughs in experiments were made several years ago, in which a
semiconductor quantum dot (QD) device was subtly designed by using nano-electronics 
technique and the conductance through this QD was measured to show the well-known Kondo 
effects, \cite{Goldhaber1,Cronenwett} the study of electronic transport through QD has 
become current interesting subject. 
\cite{Simmel,Ralph,Schmid,Sasaki,Goldhaber2,Wiel,Meir1,Meir2,Hershfield1,Ng1,Wingreen,Yeya
ti,Craco} So far, the main features of the Kondo effect that have been explored in QD are 
Kondo-assisted enhancement of conductance, its specific temperature dependence, a peak 
splitting in a magnetic field, zero-bias maximum of differential conductance in the Kondo 
regime, and integer-spin Kondo effect.

Electrical current through a conductor is always fluctuation with time and manifests the 
consequence of the quantization of the charge carriers, which is usually refered to as the 
shot noise in literature. \cite{Blanter} The noise characteristic of a conductor 
is important and interesting because the spectrum of shot noise contains some 
information related to electron transport in the conductor, which can not be 
obtained only through measuring the conductance. For example, shot noise 
experiments can determine the kinetics of electron, and reveal information on 
the correlations of electronic wave function. Therefore, shot noise has focused 
considerable investigation in mesoscopic systems. 
\cite{Buttiker,Reznikov,Kumar,Iannaccone,Kuznetsov,Nagaev,Jong,Chen,Wei,Hershfield2,Yamagu
chi,Ding} It is known that the zero-frequency shot noise $S(0)$ for a classical 
conductor is characterized by the Poisson value $S_{P}(0)=2e\langle I\rangle$ 
($\langle I\rangle$ is the average current), while the 
shot noise in a non-interacting mesoscopic conductor is always reduced by the 
Pauli exclusion in comparison with the Poisson value. Of course, shot noise is 
also influenced by electron-electron interaction. However, it seems that, depending 
on the details of the systems under investigation, both 
suppression \cite{Reznikov,Kumar} and enhancement \cite{Iannaccone,Kuznetsov} of 
shot noise from the classical value due to Coulomb interaction are observed by 
several experiments. 

Up to now, most of the theoretical studies on shot noise in mesoscopic systems are 
concentrated on the non-interacting electrons. To our knowledge, only a few papers deal 
with the shot noise of QD in the strongly correlated Kondo regime. Hershfield 
\cite{Hershfield2} 
computed perturbatively the zero-frequency current noise in the Hartree approximation 
based on the Green's function (GF) approach. He found the interaction can either enhance 
or reduce the shot noise. Yamaguchi and Kawamura \cite{Yamaguchi} performed a 
complementary analysis by treating the tunneling Hamiltonian perturbatively and revealed a 
large suppression as compared with the Poisson value due to the on-site Coulomb 
interaction. However both of their results are valid for the high-temperature regime and 
do not describe the Kondo physics because of their perturbation scheme. Ding and Ng 
\cite{Ding} calculated the zero-frequency and frequency-dependent shot noise for the Kondo 
regime at low temperature $T=T_{K}$ ($T_{K}$ denotes the Kondo temperature) by employing 
equation-of-motion method and Ng's ansatz for the correlation GF. Their results also 
demonstrated suppression of the shot noise below the non-interacting value for any applied 
voltage. They evaluated the retarded GF of the current operator and the occupation number 
operator instead of the current-current correlation GF, which must be treated for 
investigation of shot noise by means of nonequilibrium GF approach. Then shot noise 
problems in strongly correlated systems remain not to be solved.

More recently, a new slave-boson mean field (SBMF) approach has been developed to 
investigate transport through QD \cite{Dong} with arbitrary strength of the Coulomb 
interaction,  which is an extension of the saddle-point approximation to the 
auxiliary-boson functional integral method for the Anderson models, suggested by Kotliar 
and Ruckenstein, \cite{Kotliar} to the nonequilibrium situations. It has been confirmed 
that this formulation's simplest saddle-point approximation 
for all introduced Bose fields and Lagrange multipliers is, at zero temperature, 
equivalent to the results derived from the Gutzwiller variational wave function, 
\cite{Kotliar} the well-known analytical approach for strongly correlated fermions. 
It is believed that since more auxiliary parameters, which have unambiguous 
physical meanings, are introduced in this scheme than in the usual slave-boson 
formulation, this new SBMFT provides more precise description for the Kondo-type 
transport through QD. \cite{Dong} The purpose of the present paper is to further 
generalize this finite-$U$ approach to investigate the nonlinear shot noise of QD 
with and without the influence of magnetic fields.  

We arrange the rest parts of the paper as following. In the second section we briefly 
discuss the equivalent slave-boson field Hamiltonian and self-consistent equations 
to determine the unknown expectation values for slave-boson operators within the 
SBMF approach. Meanwhile, we define the ``average'' noise $S(\omega)$ and the noise 
$S_\alpha(\omega)$ of the lead $\alpha$ and directly give their expressions. At zero 
temperature, we find that the ``average'' zero-frequency shot noise coincides with the 
zero-frequency shot noise for any lead $\alpha$, but this fact is not always satisfied in 
general condition. The numerical calculation and discussion are performed in the following 
section. Fano factor which characterizes the deviation of the shot noise from the 
classical result is obtained and its limit value at low applied voltage is analysed. 
The concrete resutls clearly demonstrate that the strong on-site Coulomb interaction 
can significantly reduce the Fano factors of QD in the Kondo regime, thus largely 
suppress the shot noise below the Poisson value. On the other hand, application of 
magnetic fields enhances the
Fano factor evidently, which can be understood by the fact that the Kondo induced 
enhancement of the transmission probability through QD is weakened under the 
influence of magnetic fields. As our knowledge, this is the first time in literature
to reveal the effect of magnetic fields on the shot noise in the strongly 
correlated QD. Finally, a conclusion is given in Section 5.     

\section{Finite-$U$ Slave-Boson Mean Field Approach and Nonlinear Shot Noise Formula}

Transport through a QD coupled to two reservoirs in the presence of external voltages 
and of magnetic fields can be described by the Anderson single impurity model:
\begin{equation}
H=\sum_{\sigma k \alpha}\epsilon _{k \alpha\sigma
}'c_{k \alpha\sigma }^{\dagger }c_{k \alpha\sigma }+\sum_{\sigma
}\epsilon_{d\sigma}c_{d\sigma }^{\dagger }c_{d\sigma }+
 Un_{d\uparrow }n_{d\downarrow }+\sum_{\sigma
k \alpha}(V_{\alpha}c_{k \alpha\sigma }^{\dagger }c_{d\sigma
}+{\rm {H.c.}}),
\label{hamiltonian1}
\end{equation}
where $\epsilon _{k \alpha\sigma }'=\epsilon _{k \alpha\sigma }+V_{\alpha}$ represents 
the conduction electron energy under the application of the external voltage $V_{\alpha}$ 
to the lead $\alpha$. $c_{k \alpha\sigma}^{\dagger
}$ ($c_{k \alpha\sigma }$) are the creation (annihilation)
operators for electrons in the lead $\alpha$ ($={\rm L,R}$). When a total
external voltage $V$ is applied between the two leads, their
chemical potential difference is $\mu_{\rm L}-\mu_{\rm R}=eV$. The two leads are 
assumed to be in local equilibrium and their distribution functions are given by 
the Fermi distribution functions $f_\alpha(\omega)=[1+\exp{(\omega-\mu_\alpha)/
k_{\rm B}T}]^{-1}$ ($\alpha=$L or R). This assumption is practically correct because 
the two reservoirs respond to an applied field much faster than the center region, 
{\it i.e.}, the QD. Under the influence of a magnetic field $B$, the discrete 
energy levels $\epsilon_{d}$ in the QD are split into $\epsilon_{d\sigma}\equiv
\epsilon_{d}+\sigma h$ ($\sigma=\pm 1$) for up and down spins, where $2h=g\mu_{\rm B}B$ 
($g$ is the Land\'e factor and $\mu_{\rm B}$ the Bohr magneton) is the Zeeman energy. 
The other parameters $U$, and $V_{\alpha}$ stand for the Coulomb interaction, and the 
coupling between the QD and the reservoir $\alpha$, respectively. According to the KR 
slave-boson representation,\cite{Kotliar} we introduce four auxiliary Bose fields $e$, 
$p_\sigma$ ($\sigma=\pm 1$), and $d$, which act respectively as projection operators 
onto the empty, singly occupied (with spin up and down), and doubly occupied 
electronic states at the QD. In order to eliminate additional unphysical states, 
three constraints have to impose on these bosons
\begin{eqnarray}
&&\sum_{\sigma}p_\sigma^\dagger p_\sigma+e^\dagger e+d^\dagger d=1,
\label{con1} \\
&&c_{d\sigma}^\dagger c_{d\sigma}=p_{\sigma}^\dagger
p_\sigma+d^\dagger d,
\,\,\,\,\,\sigma=\pm 1.\label{con2}
\end{eqnarray}
Equations (\ref{con1}) and (\ref{con2}) are the completeness relation
and the condition for the correspondence between fermions and bosons, respectively. In the 
physical subspace defined by these constraints, the fermion
operators $c_{d\sigma}^\dagger$ and $c_{d\sigma}$ of the QD in the hopping terms are
replaced by
\begin{equation}
z_\sigma^\dagger c_{d\sigma}^\dagger,\qquad
c_{d\sigma}z_{\sigma},
\end{equation}
so that the matrix elements are the same in the
combined fermion-boson Hilbert space as those in the original one
Eq. (\ref{hamiltonian1}). Here
\begin{equation}
z_\sigma=\left (1-d^\dagger d-p_\sigma^\dagger p_\sigma
\right)^{-1/2} \left ( e^\dagger p_\sigma+p_{\bar {\sigma}}^\dagger d\right )
\left (1-e^\dagger e-p_{\bar {\sigma}}^\dagger
p_{\bar {\sigma}}\right )^{-1/2}. \label{z}
\end{equation} 
Therefore, the Hamiltonian
(\ref{hamiltonian1}) can be replaced by the following effective Hamiltonian in terms 
of auxiliary boson operators and of decorated fermion operators:
\begin{eqnarray}
H_{\rm eff}&=&\sum_{k \alpha\sigma}\epsilon _{k \alpha\sigma
}'c_{k \alpha\sigma }^{\dagger }c_{k \alpha\sigma }+\sum_{\sigma
}\epsilon _{d\sigma}c_{d\sigma }^{\dagger }c_{d\sigma }+
Ud^\dagger d+\sum_{k \alpha\sigma}(V_{\alpha
}c_{k \alpha\sigma }^{\dagger } c_{d\sigma }z_\sigma+{\rm
{H.c.}})
\cr &\hphantom{=}& +\lambda^{(1)}(\sum_{\sigma }p_{\sigma }^{\dagger }p_{\sigma }+
e^{\dagger}e+d^{\dagger }d-1)+
\sum_{\sigma }\lambda_{\sigma }^{(2)}(c_{d\sigma }^{\dagger }c_{d\sigma }-p_{\sigma
}^{\dagger }p_{\sigma }-d^{\dagger }d).
\label{hamiltonian2}
\end{eqnarray}
The constraints are incorporated via the three Lagrange
multipliers, $\lambda^{(1)}$ and $\lambda_\sigma^{(2)}$. Under the framework of SBMF 
approach, 
the four slave Bose fields can be assumed as c-numbers and replaced by their
corresponding expectation values. It will be seen in the following that, under 
this approximation, transport through the QD can be characterized by the seven 
factitious parameters $e$, $p_\sigma$, $d$, $\lambda^{(1)}$, and $\lambda_\sigma^{(2)}$.

Since the Hamiltonian (\ref{hamiltonian2}) describing the leads is noninteracting, the 
unperturbed retarded (advanced) GFs $g_{k\alpha\sigma}^{r,a}(t,t')$ and ``lesser'' 
(``greater'') GFs $g_{k\alpha\sigma}^{<,>}(t,t')$ for the lead $\alpha$ are
\begin{eqnarray}
g_{k\alpha \sigma}^{r(a)}(t,t')&\equiv&\mp i\theta(\pm t \mp t') \langle 
\{c_{k\alpha\sigma}^{\dagger}(t'),
c_{k\alpha\sigma}(t)\}\rangle=\mp i\theta(\pm t \mp t') 
e^{-i\epsilon_{k\alpha\sigma}(t-t')} , \label{g1}\\
g_{k\alpha \sigma}^{<}(t,t')&\equiv& i\langle c_{k\alpha\sigma}^{\dagger}(t')
c_{k\alpha\sigma}(t)\rangle = i f_{\alpha}(\epsilon_{k\alpha\sigma})e^{-i\epsilon_{k\alpha 
\sigma}(t-t')},
\label{g2} \\
g_{k\alpha \sigma}^{>}(t,t')&\equiv&-i\langle 
c_{k\alpha\sigma}(t)c_{k\alpha\sigma}^{\dagger}(t')
\rangle= -i[1- f_{\alpha}(\epsilon_{k\alpha\sigma})]e^{-i\epsilon_{k\alpha 
\sigma}(t-t')}.\label{g3}
\end{eqnarray}
However for the transport problems concerned in this paper, the electrons in the QD are 
in a nonequilibrium state, to be determined by their coupling to the two leads and to 
the applied voltage. In order to describe the nonequilibrium state of electrons, we 
define the retarded (advanced) and lesser (greater) GFs for the QD as follows: 
$G_{d\sigma}^{r(a)}(t,t')\equiv \pm i\theta (\pm t \mp t')\langle \{ c_{d\sigma}(t), 
c_{d\sigma}^{\dagger}(t') \}\rangle$, $G_{d\sigma}^{<}(t,t')\equiv i\langle   
c_{d\sigma}^{\dagger}(t')c_{d\sigma}(t) \rangle$ and $G_{d\sigma}^{<}(t,t')\equiv 
-i\langle c_{d\sigma}(t) c_{d\sigma}^{\dagger}(t') \rangle$. It is clear that for 
the effective Hamiltonian (\ref{hamiltonian2}) the Fourier transforms 
$G_{d\sigma}^{r,a,<}(\omega)$ of these GFs can be readily given, in the wide-band limit, 
as
\begin{eqnarray}
G_{d\sigma}^{r(a)}(\omega)&=&\frac
{1}{\omega-\tilde{\epsilon}_{d\sigma}\pm i \tilde{\Gamma}_{\sigma}},
\label{rGF} \\
G_{d\sigma}^<(\omega)&=&\frac{i\tilde{\Gamma}_{\sigma}[f_L(\omega)+f_R(\omega)]}{(\omega-
\tilde{\epsilon}_{d\sigma})^2+\tilde{\Gamma}_{\sigma}^2},
\label{cGF}
\end{eqnarray}
where $\tilde{\Gamma}_\sigma=(\Gamma_L+\Gamma_R)|z_\sigma|^2/2$ with 
$\Gamma_\alpha=2\pi \sum_{k \alpha} |V_\alpha|^2 \delta(\omega-\epsilon_{k \alpha 
\sigma})$ 
being the coupling strength between the QD level and the lead $\alpha$. Note that 
these formula are similar with those for noninteracting electrons, except with the 
effective energy level 
$\tilde{\epsilon}_{d\sigma}=\epsilon_{d\sigma}+\lambda_\sigma^{(2)}$ 
and the effective coupling constant $\tilde{\Gamma}_\sigma$ instead, which renormalize 
the GFs of the QD due to the strong Coulomb repulsion under the approximation employed 
here. In the present paper, we focus our attention on the symmetric systems 
$\Gamma_L=\Gamma_R=\Gamma$ and take the coupling strength $\Gamma$ as the energy unit 
throughout the paper. Moreover, the greater GF can be easily obtained from the 
relationship 
$G^{r}-G^{a}=G^{>}-G^{<}$. Besides, some other GFs are used in the following derivation 
of electric current and shot noise spectral. So we definite and treat them in advance. 
For example, we define the lesser GFs: 
$G_{d\sigma,k\alpha\sigma}^{<}(t,t')\equiv i\langle c_{k\alpha\sigma}^{\dagger}(t')
c_{d\sigma}(t)\rangle$, $G_{k\alpha\sigma,d\sigma}^{<}(t,t')\equiv i\langle 
c_{d\sigma}^{\dagger}(t')c_{k\alpha\sigma}(t)\rangle$, and 
$G_{k\alpha\sigma,k'\beta\sigma}^{<}(t,t')\equiv i\langle c_{k'\beta\sigma}^{\dagger}(t')
c_{k\alpha\sigma}(t)\rangle$ and of course their corresponding retarded, advanced, 
and greater GFs. With the effective Hamiltonian (\ref{hamiltonian2}), these GFs can be 
readily related with the GFs of the QD by applying the Langreth analytic continuation 
rules \cite{Langreth}:
\begin{eqnarray}
G_{d\sigma,k\alpha \sigma}^{<}(t,t')&=& \int dt_{1} V_{\alpha}^{*} z_\sigma^{*} 
\left [ G_{d\sigma}^{<}(t,t_{1}) g_{k\alpha \sigma}^{a}(t_{1},t') + 
G_{d\sigma}^{r}(t,t_{1}) 
g_{k\alpha \sigma}^{<}(t_{1},t')\right ],\label{gg1} \\
G_{k\alpha \sigma,d\sigma}^{<}(t,t')&=& \int dt_{1} V_{\alpha} z_\sigma 
\left [ g_{k\alpha \sigma}^{<}(t,t_{1}) G_{d\sigma}^{a}(t_{1},t') + g_{k\alpha 
\sigma}^{r}(t,t_{1}) 
G_{d\sigma}^{<}(t_{1},t')\right ], \label{gg2} \\
G_{k\alpha\sigma,k'\beta\sigma}^{<}(t,t')&=& g_{k\alpha\sigma}^{<}(t,t')\delta_{kk'}
\delta_{\alpha\beta}+\int dt_{1} V_{\beta} z_\sigma \left [ g_{k\alpha\sigma}^{<}(t,t_{1}) 
G_{d\sigma,k'\beta\sigma}^{a}(t_{1},t')+g_{k\alpha\sigma}^{r}(t,t_{1})G_{d\sigma,k'\beta
\sigma}^{<}(t_{1},t') \right ] \label{gg3}.
\end{eqnarray}

The current operator flowing from the lead $\alpha$ to the QD can be evaluated from the 
time evolution of the occupied number operator of the lead
\begin{eqnarray}
\hat{I}_\alpha(t)&=&-\frac{e}{\hbar}\Big\langle \frac{d\hat{N}_\alpha}{dt}\Big\rangle=
-i\frac{e}{\hbar}\Big [H_{\rm eff}, \sum_{k\alpha\sigma}c_{k\alpha\sigma}^{\dagger}(t)
c_{k\alpha\sigma}(t)\Big ] \cr
&=&i\frac{e}{\hbar}\sum_{k\alpha} [V_\alpha z_\sigma c_{k\alpha\sigma}^\dagger(t) 
c_{d\sigma}(t)-V_\alpha^{*} z_\sigma^{*} c_{d\sigma}^\dagger(t) c_{k\alpha\sigma}(t)]. 
\label{i}
\end{eqnarray}
Because the electric current fluctuates, the currents flowing into QD and flowing out of 
QD are not in balance. The terminal current is given by the average current 
$\hat{I}=(\hat{I}_L-\hat{I}_R)/2$. Its statistical expectation yields the current 
through QD we are interested in. By the help of the Dyson equations (\ref{gg1}) and 
(\ref{gg2}),
the current formula through the QD takes the form \cite{Dong}
\begin{equation}
I=\frac{{\rm e}}{\hbar}\sum_{\sigma }\int
d\omega \tilde{\Gamma}_\sigma
\left\{ f_{L}(\omega )-f_{R}(\omega )\right\}
\rho _{\sigma }(\omega ),
\label{current}
\end{equation}
where $\rho_\sigma=-(1/\pi) {\rm Im} G_{d\sigma}^r (\omega)$ is the spectral 
density-of-state (DOS) of the electron in the QD. Starting from the constraints 
(\ref{con1}), (\ref{con2}) and the equation of motion of the slave-boson operators 
from the Hamiltonian (\ref{hamiltonian2}), we can yield the following self-consistent set 
of equations within the SBMF approach as ($\sigma=\pm 1$): \cite{Dong} 
\begin{eqnarray}
&&\sum_{\sigma }|p_{\sigma }|^{2}+|e|^{2}+|d|^{2}=1,  \label{set1}
\\
&&\frac{1}{2\pi i}\int d\omega G_{d\sigma
}^{<}(\omega )=|p_{\sigma }|^{2}+|d|^{2},  \label{set2c} \\
&&\frac{1}{2\pi i}\sum_{\sigma }\frac{\partial \ln z_{\sigma }}{
\partial e}\int d\omega G_{d\sigma }^{<}(\omega )\left( \omega -\tilde{
\epsilon}_{d\sigma }\right)+2\lambda ^{(1)}e =0, \\
&& \frac{1}
{2\pi i}\sum_{\sigma' }\left(\frac{\partial \ln z_{\sigma' }}{\partial 
p_{\sigma}^\dagger}+
\frac{\partial \ln z_{\sigma' }}{\partial p_{\sigma}}\right) \int d\omega
G_{d\sigma' }^{<}(\omega )
\left( \omega -\tilde{\epsilon}_{d\sigma' }\right)+2\left( \lambda ^{(1)}-\lambda _{\sigma 
}^{(2)}\right) p_{\sigma } 
=0, \\
&&\frac{1}{2\pi i}\sum_{\sigma }\frac{\partial \ln z_{\sigma }}{\partial d}\int
d\omega G_{d\sigma }^{<}(\omega )\left( \omega -\tilde{\epsilon}_{d\sigma
}\right)+
2\left( U+\lambda ^{(1)}-\sum_{\sigma }\lambda _{\sigma }^{(2)}\right) d =0.  
\label{set5c}
\end{eqnarray}
Therefore, these equations (\ref{set1})-(\ref{set5c}) form a closed set of self-consistent 
equations, which 
can define the seven parameters, and thus describe linear and nonlinear transport 
through the QD under finite external voltages and magnetic fields. 

The shot noise is resulted from the quantization of the charge and to observe it 
we have to investigate the nonequilibrium (transport) state of the systems. While, 
from the point of view of practical experiments, it is the external voltage 
fluctuations which are actually measured and which eventually are converted to 
current fluctuations. As a result of this fact, the noise can be taken as a 
response of current to a small amplitude high-frequency voltage superposed on a 
dc bias: $V_{\alpha}(t)=V_{\alpha 0}+V_{\alpha 1}\cos(\omega t)$, with $V_{\alpha 0}$ 
and the $V_{\alpha 1}$ term denote the dc and ac bias, respectively. As well the 
noise power spectrum turns out to be related to the ac conductance in the presence 
of this high-frequency voltage $V_{\alpha}(t)$. In the linear transport region 
$V_{\alpha 0}\rightarrow 0$, this gives rise to the well-known Einstein relation 
$S(\omega)=4kT\sigma(\omega)$ ($\sigma(\omega)$ is the small signal ac conductance). 
Naturally, we can use the Hamiltonian (\ref{hamiltonian1}) to describe the system 
under the influence of the time-alternating voltage $V_{\alpha}(t)$ and further 
calculate transport and noise for the QD. It is obvious that after a transient 
process the system will be driven to an oscillatory steady state, in which the 
physics quantities will perform as a sum of a dc part under the influence of 
$V_{\alpha 0}$ and a small amplitude oscillation part at the single driving 
frequency $\omega$ (fluctuation part in the present problem), if the high-frequency 
bias is sufficiently small. For example, the current operator $\hat{I}_{\alpha}(t)$ 
now consists of both the average current $\langle {I}_{\alpha}(t)\rangle$ and a 
small fluctuation $\delta \hat{I}_{\alpha} (t)$
\begin{eqnarray}
\hat{I}_{\alpha}(t)&=&\langle {I}_{\alpha}(t)\rangle+\delta \hat{I}_{\alpha}(t) \cr
&=& I_{\alpha}+\delta \hat{I}_{\alpha}(t) \label{di}
\end{eqnarray}
with
\begin{equation}
\langle \delta \hat{I}_{\alpha}(t)\rangle=0.
\end{equation}
With no reason, an important assumption can be made as follows: those introduced 
auxiliary Bose fields and the Lagrange multipliers also obey this discipline 
within SBMF approach:
\begin{eqnarray}
\hat{\cal O}(t)&=&{\cal O}_{0}+\delta \hat{\cal O}(t),\\
\lambda (t)&=&\lambda_{0}+\delta \lambda (t)
\end{eqnarray}
with
\begin{eqnarray}
\langle \delta \hat{\cal O}(t)\rangle&=& 0,\\
\langle \delta \lambda (t)\rangle &=&0,
\end{eqnarray}
where the operator $\hat{\cal O}$ represents any one of the slave-boson operators 
$e$, $p_{\sigma}$, and $d$ and ${\cal O}_{0}$ the corresponding expectation value. 
$\lambda$ denotes the Lagrange multipliers $\lambda^{(1)}$ and $\lambda_{\sigma}^{(2)}$ 
and $\lambda_{0}$ the constant value. Owing to the sufficiently small amplitudes 
of these fluctuations of Bose fields, we can ignore their minor contribution and 
only remain their zero order terms, the expectation values ${\cal O}_{0}$ and 
$\lambda_{0}$, in the calculation for the correlation of the current fluctuation. 
This is our central presumption in this paper. Discussion about the validity of 
this approximation is beyond the scope of the present paper and the effects of 
fluctuation of these Bose fields on transport will be investigated in our future 
publication. With this consideration, the effective Hamiltonian (\ref{hamiltonian2}) 
can still be utilized to depict transport through the QD under the voltage 
$V_{\alpha}(t)$ but with the constant expectation values ${\cal O}_{0}$ and 
$\lambda_{0}$ instead. It turn out to be that this procedure can provide a 
considerable precise prescription for kondo-type transport through QD with and 
without magnetic fields. \cite{Dong} In the following, starting from this 
Hamiltonian (\ref{hamiltonian2}), we will investigate the shot noise of QD in the 
Kondo regime. The approximation of slave-boson mean field deals with the on-site 
Coulomb interaction as a renormalization hopping factor $z_{\sigma}$ [Eq.(\ref{z})] 
and makes the Hamiltonian for the QD system reduce to non-interacting one will 
largely simplify the calculation for the current-current correlation function.

The power density $S(\omega)$ of the tunneling current fluctuation is defined as 
the Fourier transform of the following correlation function:
\begin{equation}
S(t-t')=\langle \delta \hat{I}(t) \delta \hat{I}(t')+\delta \hat{I}(t') \delta 
\hat{I}(t)\rangle. \label{sa}
\end{equation}
Note that the current operator involved in the definition is the average current. 
So we term this noise current as ``average'' noise. However, in the practical 
experiments, one has to choose either ``L'' lead or ``R'' lead to measure the current 
and noise spectrum. Therefore, it is necessary to define the noise $S_{\alpha}(t-t')$ 
of the lead $\alpha$:
\begin{equation}
S_{\alpha}(t-t')=\langle \delta \hat{I}_{\alpha}(t) \delta \hat{I}_{\alpha}(t')+
\delta \hat{I}_{\alpha}(t') \delta \hat{I}_{\alpha}(t)\rangle. \label{sab}
\end{equation}
In order to evaluate the noise, we substitute the current operator Eq.(\ref{i}) into 
Eqs.(\ref{sa}) and (\ref{sab}) and express these quantum statistical (nonequilibrium) 
average in terms of the nonequilibrium GFs. After a length but straightforward derivation, 
we obtain:
\begin{eqnarray}
S_{\alpha}(\omega)&=&-\frac{e^2}{\hbar} \sum_{\sigma}\tilde{\Gamma}_{\sigma}^2 \int 
{d\omega_1 \over 2\pi} \big\{\left \{ f_{\alpha}(\omega_1)\left [ 
1-f_{\alpha}(\omega+\omega_1)
\right]\left[ G_{d\sigma}^{r}(\omega_1)G_{d\sigma}^{r}(\omega+\omega_1)+
G_{d\sigma}^{a}(\omega_1)G_{d
\sigma}^{a}(\omega+\omega_1)\right] \right.\cr
&&+\left[1-f_{\alpha}(\omega+\omega_1)\right]G_{d\sigma}^{<}(\omega_1)
\left[G_{d\sigma}^{r}(\omega+\omega_1)-G_{d\sigma}^{a}(\omega+\omega_1)\right]+
f_{\alpha}(\omega_1)\left[
G_{d\sigma}^{a}(\omega_1)-G_{d\sigma}^{r}(\omega_1)\right] 
G_{d\sigma}^{>}(\omega+\omega_1) \cr
&&\left.\left.-G_{d\sigma}^{<}(\omega_1)G_{d\sigma}^{>}(\omega+\omega_1)+
{i\over \tilde{\Gamma}_{\sigma}}\left\{\left[ 1-f_{\alpha}(\omega+\omega_1) \right]
G_{d\sigma}^{<}(\omega_1)-f_{\alpha}(\omega)G_{d\sigma}^{>}(\omega+\omega_1) 
\right \} \right \}+\left\{ \omega\rightarrow-\omega \right\} \right \}, \label{nsa} \\
S(\omega)&=&-\frac{e^2}{4\hbar}\sum_{\sigma} \tilde{\Gamma}_{\sigma}^2 \int {d\omega_1 
\over 2\pi}
 \left \{ \left \{ \left [
f_{L}(\omega_1)[f_{R}(\omega+\omega_1)-f_{L}(\omega+\omega_1)]+f_{R}(\omega_1)[
f_{L}(\omega+\omega_1)-f_{R}(\omega+\omega_1)]\right ]\right. \right. \cr
&&\hspace{2.5cm}\times \left[ 
G_{d\sigma}^{r}(\omega_1)G_{d\sigma}^{r}(\omega+\omega_1)+G_{d\sigma}^{a}(\omega_1)
G_{d\sigma}^{a}(\omega+\omega_1)\right] \cr
&&\left. \left. +{i \over \tilde{\Gamma}_{\sigma}}\left \{
\left[ 
2-f_{L}(\omega+\omega_1)-f_{R}(\omega+\omega_1)\right]G_{d\sigma}^{<}(\omega_1)-\left[
f_{L}(\omega+\omega_1)+f_{R}(\omega+\omega_1)\right ]
G_{d\sigma}^{>}(\omega_1)\right \}\right \}+\left \{ \omega\rightarrow-\omega\right\} 
\right \}. \label{ns}
\end{eqnarray}
It is very clear that the average noise spectrum $S(\omega)$ is different from the 
noise spectrum $S_{\alpha}(\omega)$ of the lead $\alpha$. In the nonequilibrium 
condition, the difference of the noise spectrum between the $L$ and the $R$ leads 
is also distinct. 

In this paper, we focus our attention on the zero-frequency noise spectrum. 
Assuming $\omega=0$ in the formula (\ref{nsa}) and (\ref{ns}), it is found that 
the averge zero-frequency noise $S(0)$ happens to be equal to the zero-frequency 
noise $S_{\alpha}(0)$ of the lead $\alpha$,
\begin{eqnarray}
S(0)=S_{\alpha}(0)&=&{e^2 \over 2\hbar}\sum_{\sigma} \tilde{\Gamma}_{\sigma}\int {d\omega 
\over 2\pi}
\left\{ \left[f_{L}(\omega)-f_{R}(\omega)\right]^2 
\left[G_{d\sigma}^{r}(\omega)-G_{d\sigma}^{a}(\omega)
\right]^2 \right.\cr
&&\hspace{2cm}\left.+{2i\over \tilde{\Gamma}_{\sigma}} \left\{ f_{L}(\omega)
\left[1-f_{R}(\omega) \right] +f_{R}(\omega) \left[ 1-f_{L}(\omega)\right ]\right\} 
\left[ G_{d\sigma}^{r}(\omega)-G_{d\sigma}^{a}(\omega)\right]\right\}. \label{so}
\end{eqnarray}
The zero-frequency noise power spectrum formula is exactly the same as that derived 
from the scattering matrix theory for the non-interacting electron systems, 
\cite{Blanter,Wei} except the coupling strength $\Gamma$ and GFs 
$G_{d\sigma}^{r(a)}(\omega)$ 
for the QD are replaced with the renormalized ones. It is worth noting that at 
zero temperature the equilibrium noise spectrum for the case of zero frequency is exactly 
zero, {\it i.e.}, only 
the nonequilibrium or shot noise contributs to the zero-frequency noise power spectrum. 

\section{Calculation and Discussion}

In this section, we numerically investigate the Kondo correlation effect on the 
zero-frequency noise properties of QD in the presence or absence of magnetic fields. 
Calculation is carried out only at zero temperature in the present paper. For the 
sake of simplification, an assumption that a symmetric voltage drop, $\mu_L=-\mu_R=eV/2$, 
through the whole systems is made in our calculation. Then, considering symmetric 
coupling for two tunnel barriers, the current and the zero-frequency noise spectrum 
are anti-symmetric under bias reversal.

Thanks to the simple form of the retarded and advanced GFs $G_{d\sigma}^{r(a)}(\omega)$ 
(\ref{rGF}), the integrals in Eqs.(\ref{current}) and (\ref{so}) can be done exactly 
at zero temperature. We have
\begin{eqnarray}
I&=&\frac{e}{h}\sum_{\sigma}\tilde{\Gamma}_{\sigma}\left( \arctan 
\frac{\phi-\tilde{\epsilon}_{d\sigma}}
{\tilde{\Gamma}_{\sigma}} + \arctan 
\frac{\phi+\tilde{\epsilon}_{d\sigma}}{\tilde{\Gamma}_{\sigma}}\right),\\
S(0)&=&\frac{e^2}{h}\sum_{\sigma}\tilde{\Gamma}_{\sigma}\left \{ \arctan 
\frac{\phi-\tilde{\epsilon}_{d\sigma}}{\tilde{\Gamma}_{\sigma}}+
\arctan \frac{\phi+\tilde{\epsilon}_{d\sigma}}{\tilde{\Gamma}_{\sigma}}-
\tilde{\Gamma}_{\sigma}
\left[\frac{\phi-\tilde{\epsilon}_{d\sigma}}{(\phi-\tilde{\epsilon}_{d\sigma})^2+
\tilde{\Gamma}_{\sigma}^2}+\frac{\phi+\tilde{\epsilon}_{d\sigma}}{(\phi+
\tilde{\epsilon}_{d\sigma})^2+\tilde{\Gamma}_{\sigma}^2}\right]\right\},
\end{eqnarray}
in which $\phi=eV$ is the total voltage drop between the left and right leads. It 
is clear that zero-temperature shot noise is always suppressed in comparison with 
the Poisson value $S_{P}=2eI$. What magnitude of the suppression below the Poissonian 
limit is one of the aspects of noise in mesoscopic systems which triggered many 
of the theoretical and experimental works. \cite{Blanter} The Fano factor $\gamma$ 
that is the ratio of the actual shot noise to the Poisson noise provides a useful 
device to address this {\it sub-Poissonian shot noise},
\begin{equation}
\gamma=\frac{S(0)}{2eI}. \label{Fanof}
\end{equation}
Before devoting to the numerical analysis of $\gamma$, two limit cases are 
inspected. First, it is easy to gain that in the limit of enough large voltage 
$V\rightarrow \infty$, the Fano factor $\gamma\rightarrow 0.5$ for the symmetric 
systems, \cite{Chen} which has been observed in experiment, \cite{Iannaccone} 
regardless of whether considering the strong Coulomb interaction or not. Secondly, 
in the opposite limit when the voltage difference is very small $V\rightarrow 0$, 
we have
\begin{equation}
\lim_{V\rightarrow 0}\gamma=\sum_{\sigma} \frac{\tilde{\Gamma}_{\sigma}^2 
\tilde{\epsilon}_{d\sigma}^2}{\left ( \tilde{\epsilon}_{d\sigma}^2+ 
\tilde{\Gamma}_{\sigma}^2 \right )} \Big / \sum_{\sigma} \frac{\tilde{\Gamma}_{\sigma}^2} 
{\tilde{\epsilon}_{d\sigma}^2+\tilde{\Gamma}_{\sigma}^2}.
\end{equation}
In absence of magnetic fields, it reduces to
\begin{equation}
\lim_{V\rightarrow 0}\gamma=\frac{\tilde{\epsilon}_{d}^2}{\tilde{\epsilon}_{d}^2+ 
\tilde{\Gamma}^2}.
\end{equation}

In Figs.\,1(a) and (b) we plot, respectively, $I$-$V$ characteristic and the bias 
voltage-dependent zero-frequency shot noise $S(0)$ for the QD with $U=7$ and several 
different energy levels in the Kondo regime $\epsilon_d=-1$, $-1.5$, $-2$, $-3.5$, and 
$\epsilon_d=-6$. For these parameters, the Kondo temperatures $T_{K}$ are about $0.44$, 
$0.23$, $0.14$, $0.076$, and $0.44$, respectively [the exact Bethe-ansatz give this 
dynamic energy scale $T_{K}=U\sqrt{\beta}\exp(-\pi/\beta)/2\pi$, 
$\beta=-2U\Gamma/\epsilon_{d}(U+\epsilon_{d})$]. The corresponding differential 
conductance $dI/dV$ and zero-frequency differential shot noise $dS(0)/dV$ are also 
depicted as functions of the bias voltage in Figs.\,1(c) and (d). It is clear that the 
zero-frequency shot noise is smaller than the Poisson value $S_{P}$ at the whole range of 
voltages, showing suppression of shot noise spectrum. A zero-bias maximum behavior in 
$dI/dV$-$V$ is found in Fig.\,1(c), demonstrating the typical Kondo feature for these 
systems under consideration. While the differential shot noise spectrum exhibits different 
behavior, nonzero-bias maximum, for all of these systems. 

Furthermore, for the sake of comparison, we plot the zero-frequency shot noise spectrum 
and the differential shot noise for the QD without on-site Coulomb interaction $U=0$ in 
Fig.\,2 and its inset. 
Evidently we can observe from Fig.\,1(b) and Fig.\,2, that for these system parameters 
$\epsilon_d=-1$, $-1.5$, and $-2$, the zero-frequency noise power pectrum $S(0)$ in the 
finite interacting case is significantly smaller than those of the non-interacting 
systems. In equilibrium, the resonance is reached only for the energy level $\epsilon_d$ 
of the non-interacting QD aligning with the chemical potentials $\mu_{L(R)}$ of the two 
reservoirs, {\it i.e.}, $\epsilon_d=0$ [we assume $\mu_{L(R)}=0$ in calculation, see 
Fig.\,3(c)]. If the energy level $\epsilon_d$ of the QD is far below the chemical 
potentials of the two reservoirs the transmission probability $T$ is much less than one, 
\cite{Dong} which means a large suppression of current through the device. But the strong 
on-site Coulomb interaction can keep the resonance up to $\epsilon_d+U=0$. This is the 
well-known Kondo effect that has been observed recently for QD in experiments. 
\cite{Goldhaber1,Cronenwett,Simmel,Ralph,Schmid,Sasaki,Goldhaber2,Wiel} Therefore, in the 
whole range from $\epsilon_d=0$ to $\epsilon_d=-U$ the transmission probability through 
the QD remains to be nearly $1$. As a consequence, the properties of shot noise are 
complicated in the presence of strong Coulomb interaction and a universal presentation 
cannot be summarized in this situation.

A convenient way of addressing how the Coulomb interaction affects the shot noise power
spectrum in the Kondo regime is to explore the ratio between the shot 
noise and the current, {\it i.e.}, the Fano factor $\gamma$. In Figs.\,3, we depict 
the Fano factor with (a) and without (b) the on-site Coulomb repulsion as a function 
of the external voltage for the same system parameters as those in Fig. 1. As expected, 
the Fano factors approach to $0.5$ at large voltage for both of the two cases. However, 
excepting this limit behavior, tremendous difference between the two cases is 
explicitly observed in the whole range of voltages under consideration. We will interpret 
these phenomena in a qualitative way as follows.   

The classical noise theory yields that the noise $S(0)\propto T(1-T)$, which completely 
vanishes for a perfectly transparent scatterer $T=1$ and for a perfectly reflecting 
scatterer $T=0$ and reaches a maximum at $T=1/2$. \cite{Blanter} Meanwhile, the 
Landauer-B\"uttick scattering theory gives the current to be proportional to 
the transmission coefficient $I\propto T$. Thus we have $\gamma\propto (1-T)$. In 
the limit of low-transparency $T\ll 1$ the shot noise $S(0)$ is given by the Poisson 
value $S_{P}$ and the Fano factor $\gamma$ is equal to $1$. So it is helpful to 
review the transmission coefficient through QD. \cite{Ng2,Delft} In nonequilibrium 
condition and under the influence of a magnetic field, the transmission amplitude 
is related to the retarded GF $G_{d\sigma}^{r}(\omega)$ of the QD, 
\begin{equation}   
t_{d}=\frac{1}{2}\sum_{\sigma}\tilde{\Gamma}_{\sigma}G_{d\sigma}^{r}(\phi), \label{tram}
\end{equation}
at zero temperature. Substituting the expression of the retarded GF Eq.(\ref{rGF}) 
into Eq.(\ref{tram}), we derive the transmission coefficient $|t_{d}|^2$ in the linear 
limit,
\begin{equation}
|t_{d}|^2=\frac{1}{4}\sin^2 \left(\pi n_{\sigma} \right) \sin^2 \left(\pi n_{\bar{\sigma}} 
\right) 
\left\{ 4+\left[ \cot\left(\pi n_{\sigma} \right) +\cot\left(\pi n_{\bar{\sigma}} 
\right)\right]^2\right\}, \label{tranc1}
\end{equation}
where $n_{\sigma}$ denotes the occupation number of electrons with spin $\sigma$ in QD, 
\cite{Dong}
\begin{equation}
n_{\sigma}=\frac{1}{2}-\frac{1}{\pi}\arctan \left( 
\frac{\tilde{\epsilon}_{d\sigma}}{\tilde{\Gamma}_{\sigma}} \right).
\end{equation}
In absence of a magnetic field, the transmission coefficient $|t_{d}|$ 
Eq.(\ref{tranc1}) reduces to the formula of the Ref.(\onlinecite{Delft}) (Eq.(3) 
in that paper),
\begin{equation}
|t_d|^2=\sin^2 \left(\pi n_{\sigma} \right). \label{tranc2}
\end{equation}
Using Eq.(\ref{tranc2}), we calculate the transmission coefficient $|t_{d}|^2$ 
of the QD under several applied external voltages with and without considering the 
Coulomb interaction, which are plotted as a function of the energy level in Fig.\,3(c). 
It is obvious that for those energy levels, $|t_{d}|^2$ of the 
non-interacting QD are always smaller than those of the strongly interacting QD 
in zero voltage limit. Specially, we can easily observe that an approximatively 
full transparency is established in the vicinity of the electron-hole symmetry 
$\epsilon_d=-U/2=-3.5$. 
At this point, the Kondo effect is of most importance and the transmission 
probability $|t_{d}|^2$ through the QD is nearly equal to $1$. Thus it is rational 
that the Fano factor demonstrates the most pronounced reduction and an approximate 
zero value at small voltages, as shown by the dot-dashed line in Fig.\,3(a). Therefore it 
is self-evident from the formula $\gamma\propto 1-|t_{d}|^2$ that the Coulomb 
interaction suppresses the Fano factors in the Kondo regime in comparison with the 
results without the Coulomb interaction. Moreover, the external voltages that reduce 
the transmission probability evidently in the interacting QD can cause the Fano 
factor increasing monotonically. But in the opposite non-interaction case the Fano factor 
displays a profound behavior which is dependent on the energy level. In addition, due to 
the symmetric structure studied here, the energy-level-dependent transmission 
probability is exactly symmetric about the electron-hole symmetry. This is why 
the Fano factor for $\epsilon_{d}=-6$ coincides with that for $\epsilon_{d}=-1$. 
In contrast, because of the low-transparency for the non-interacting QD with 
$\epsilon_{d}=-6$, the Fano factor indicates a weak voltage-dependence and it is 
nearly equal to $1$. In summary, we can claim that the electron-electron interaction 
can largely suppress the Fano factor, i.e., the shot noise of QD systems in comparison 
with the Poisson value.  

Magnetic field is a helpful tool to probe the properties of noise. Recently, a magnetic 
field induced enhancement of noise is reported in the experiment. \cite{Kuznetsov} 
The theoretical investigation on noise under the influence of magnetic fields is 
scarce. Here we numerically study this problem in the QD. In Figs.\,4, we show the 
calculated current (a) and the zero-frequency shot noise spectrum (b) of the QD with $U=4$ 
and $\epsilon_d=-1$ versus voltages in the presence of magnetic fields from $h=0$ up to 
$0.5$. For the specially chosen parameters, the Kondo temperature $T_{K}$ of the QD is 
about $0.3$. As shown in the inset figure of Fig.\,4(a), the single Kondo peak in the 
differential conductance gradually splits with increasing magnetic fields and a 
nonzero-bias maximum appears at $h=0.5$. On the contrary, the magnetic field-dependent 
zero-frequency differential shot noise experiences a crossover from nonzero-bias maximum 
to zero-bias maximum when the magnetic field increases. Again, we find that the 
differential shot noise demonstrates very different behavior from the differential 
conductance. Therefore, measuring the shot noise spectrum can improve our understanding of 
electric properties in QD systems.         

As revealed in the experiment, \cite{Kuznetsov} the magnetic fields indeed enhance 
the Fano factor. We can observe this behavior more clearly in Fig.\,5, in 
which the Fano factors are illustrated as a function of voltages. 
This enhancement can be interpreted as the fact that the reduction of the Kondo-enhanced 
DOS due to applying of magnetic fields in QD decrease the conductance or the 
transmission probability in the Kondo regime, 
\cite{Goldhaber1,Cronenwett,Ralph,Schmid,Sasaki,Meir2,Dong} 
as is shown in the inset figure of Fig.\,5 which depicts the magnetic-field-dependent 
transmission probability for the QD at several voltages. Of course, more strong 
magnetic fields can result in more pronounced reduction of DOS in the QD, thus 
can enhance the Fano factor more significantly. 

\section{Conclusion}

We have studied the shot noise properties through QD on the basis 
of a new finite-$U$ SBMF approach and the nonequilibrium Green's function approach. 
The advantage of the present method, which the strongly interacting Hamiltonian 
for QD is transformed to a non-interacting one by introducing several auxiliary 
Boson field operators, is its capability of eliminating the crucial difficulty in 
calculating the 
current-current correlation GFs involved in the investigation of the shot noise. 
This renormalized Hamiltonian can make us deal with the single-particle GFs only 
instead of the two-particle correlation GFs. Finally, by assuming that the fluctuations 
of those slave-boson operators are neglected in the calculation, we derive an explicit 
expression for the shot noise power spectrum in terms of the Keldysh GFs in QD, 
which is similar with that for the non-interacting systems. The validity about 
this approximation and the resulting effects of including these fluctuations will 
be investigated in a separate publication. Generally speaking, the defined ``average'' 
shot noise spectrum $S(\omega)$ is different from the shot noise $S_{\alpha}(\omega)$ 
of the lead $\alpha$. Worth noting, in the limit of zero frequency the average shot 
noise coincides by chance with the noise of any leads at zero temperature.    

In the wide band limit, the integral is performed for the zero-frequency shot noise 
at zero temperature. This concrete expression is utilized to analyze the Fano factors 
in the small voltage and large voltage limits. The Fano factor $\gamma$ is a helpful 
tool to measure the enhancement or suppression of the shot noise to its Poisson value. 
We find that the property $\gamma=1/2$ for symmetric systems is universal in the 
limit of $V\rightarrow \infty$ no matter what the strong Coulomb interaction is 
presence or absence and no matter what the magnetic field is applied or not. 

This expression is also employed to carry out calculation for the zero-frequency 
shot noise power spectrum $S(0)$, zero-frequency differential shot noise $dS(0)/dV$ and 
the Fano factor $\gamma$ versus the external 
voltage at zero temperature. The numerical results show that the zero-frequency 
differential shot noise spectrum shows an obviously different behavior from the 
differential conductance, providing a beneficial information to improve understanding of 
electric properties in QD. Furthermore, our calculation reveals that the Fano factors are 
largely reduced by the Coulomb interaction, which indicates that the electron-electron 
interaction provides another mechanism, besides the Pauli exclusion rule, to further 
suppress the shot noise in QD below the Poission value.

On the other hand, the effects of magnetic fields on shot noise are also investigated 
numerically. An enhancement of the Fano factor is found due to the magnetic-field-induced 
reduction of the Kondo-enhanced DOS and the transmission probability, which is accordance 
with the physical expectation. 


\section*{Acknowledgements}

One of the authors, B. Dong, is very pleasure to acknowledge fruitful discussion 
with Dr. G. H. Ding. This work was supported by the National Natural Science Foundation
of China, the Special Funds for Major State Basic Research 
Project (grant No. 2000683), the Ministry of Science and Technology of China, the
Shanghai Municipal Commission of Science and Technology, and the
Shanghai Foundation for Research and Development of Applied
Materials.


\section*{Figure Captions}

\begin{itemize}
\item[{\bf Fig.\,1}] (a) The current $I$, (b) the zero-frequency shot noise power spectrum 
$S(0)$, (c) the differential conductance $dI/dV$, and (d) the zero-frequency differential 
shot noise spectrum $dS(0)/dV$, at zero temperature, as a function of voltage for the QD 
energy levels $\epsilon_d=-1$, $-1.5$, $-2$, $-3.5$, and $-6$ with the on-site Coulomb 
interaction $U=7$.

\vspace{1mm}

\item[{\bf Fig.\,2}] The zero-frequency shot noise power spectrum $S(0)$ at zero 
temperature as a function of voltage for the QD with $U=0$ and the same energy levels as 
Fig.\,1. The inset figure: the corresponding zero-frequency differential shot noise.    

\vspace{1mm}

\item[{\bf Fig.\,3}] (a) The Fano factor vs voltage for the QD with $U=7$. (b) The 
Fano factor vs voltage for the QD with $U=0$. The other parameters are the same 
as in Fig.\,1. (c) The transmission coefficient vs the energy level of the QD with 
$U=7$ (thick lines) and $U=0$ (thin lines) at several voltages $V=0$, $0.1$, $0.2$, 
and $0.5$ ($U=0$ only).

\vspace{1mm}

\item[{\bf Fig.\,4}] (a) The current and (b) the zero-frequency shot noise power spectrum 
at zero temperature vs voltage for the QD with the energy level $\epsilon_d=-1$ and with 
the on-site Coulomb interaction $U=4$ in different applied magnetic fields $h=0$, $0.1$, 
$0.2$, $0.3$, $0.4$, and $0.5$. Inset: (a) the differential conductance vs voltage; (b) 
the zero-frequency differential shot noise vs voltage.   

\vspace{1mm}

\item[{\bf Fig.\,5}] The Fano factor vs voltage for the QD with the same parameters 
as in Fig.\,4. Inset: The transmission coefficient vs the magnetic field at different 
voltages $V=0$, $0.2$, and $0.3$.

\end{itemize}

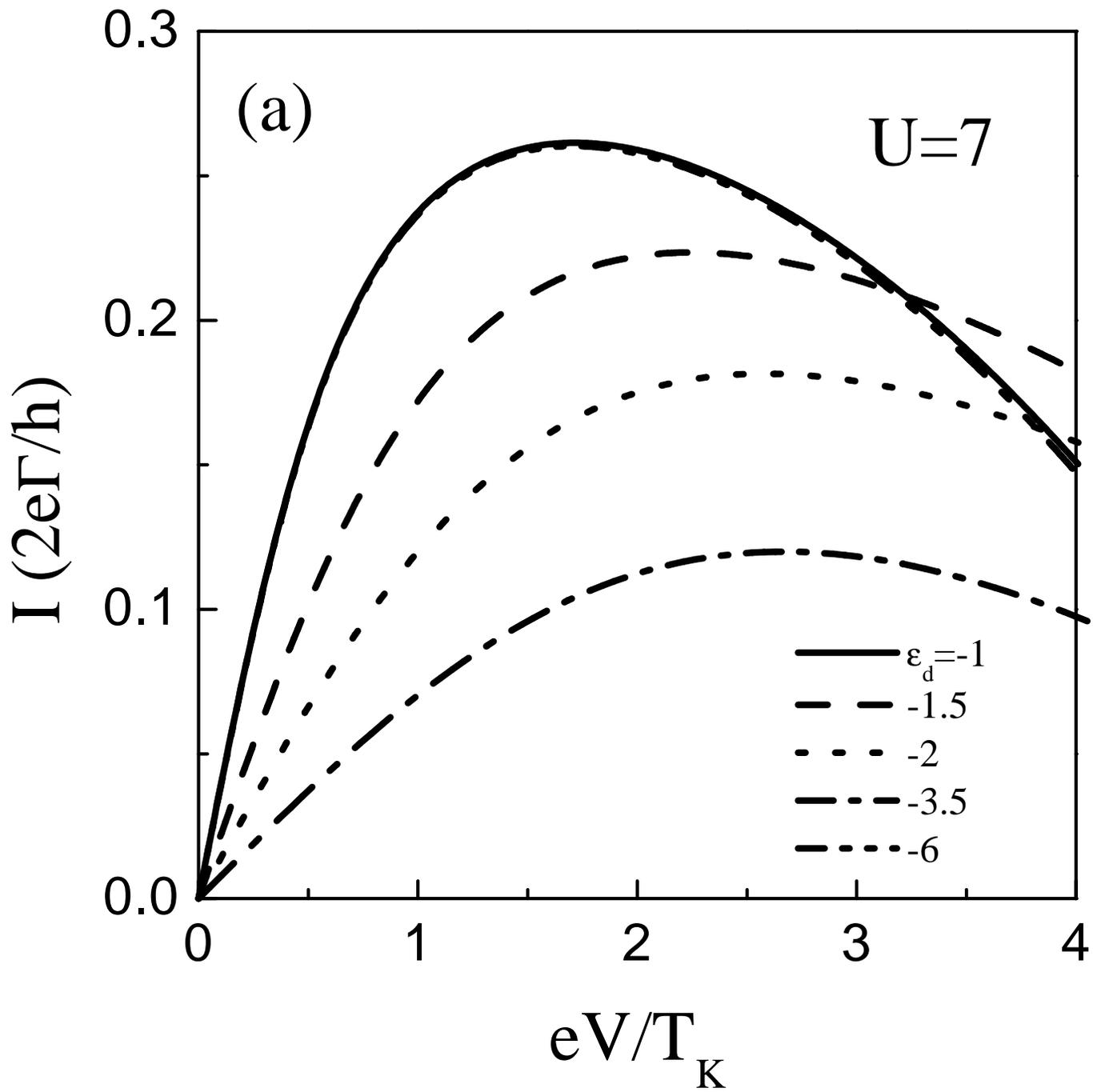
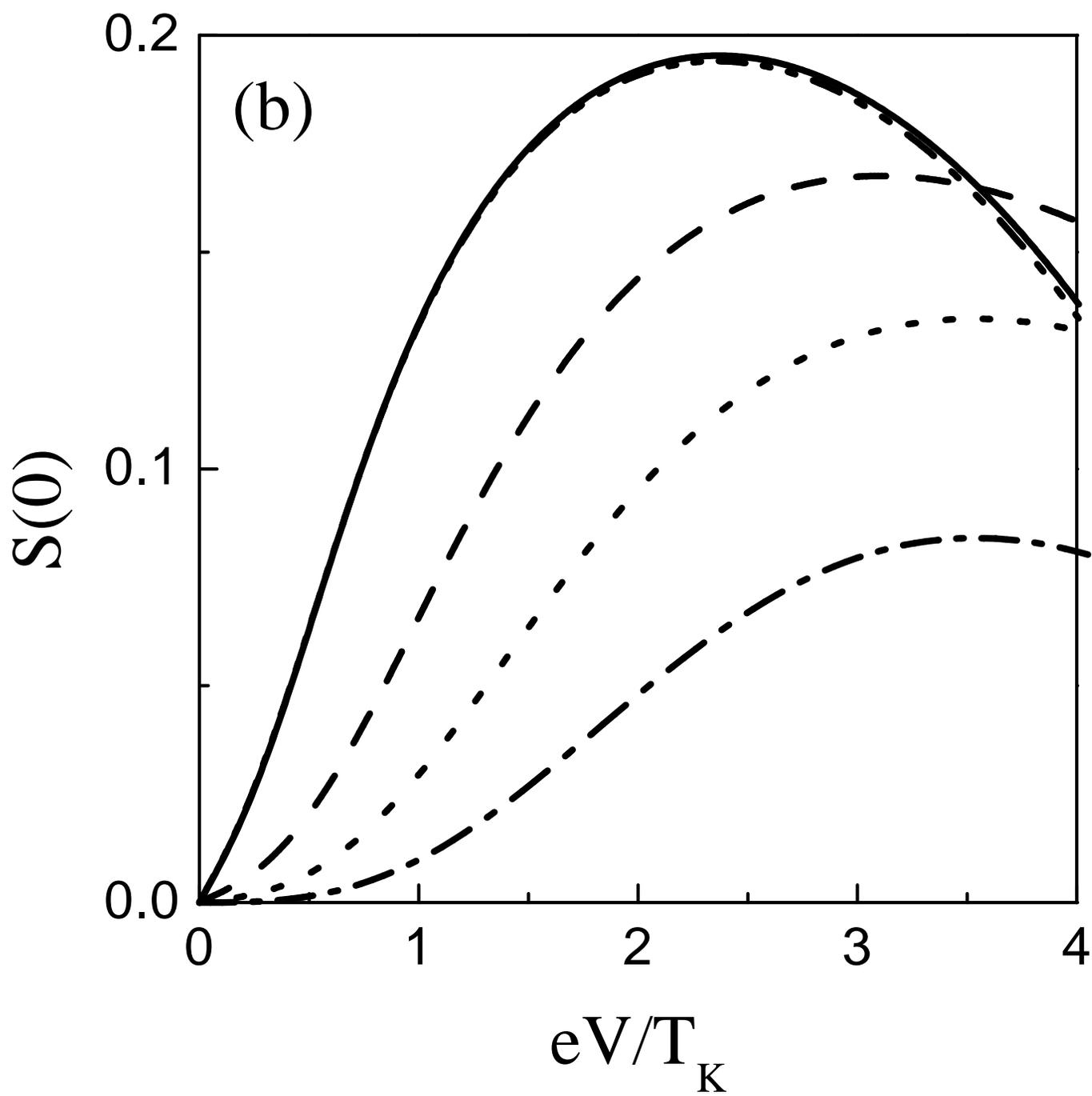
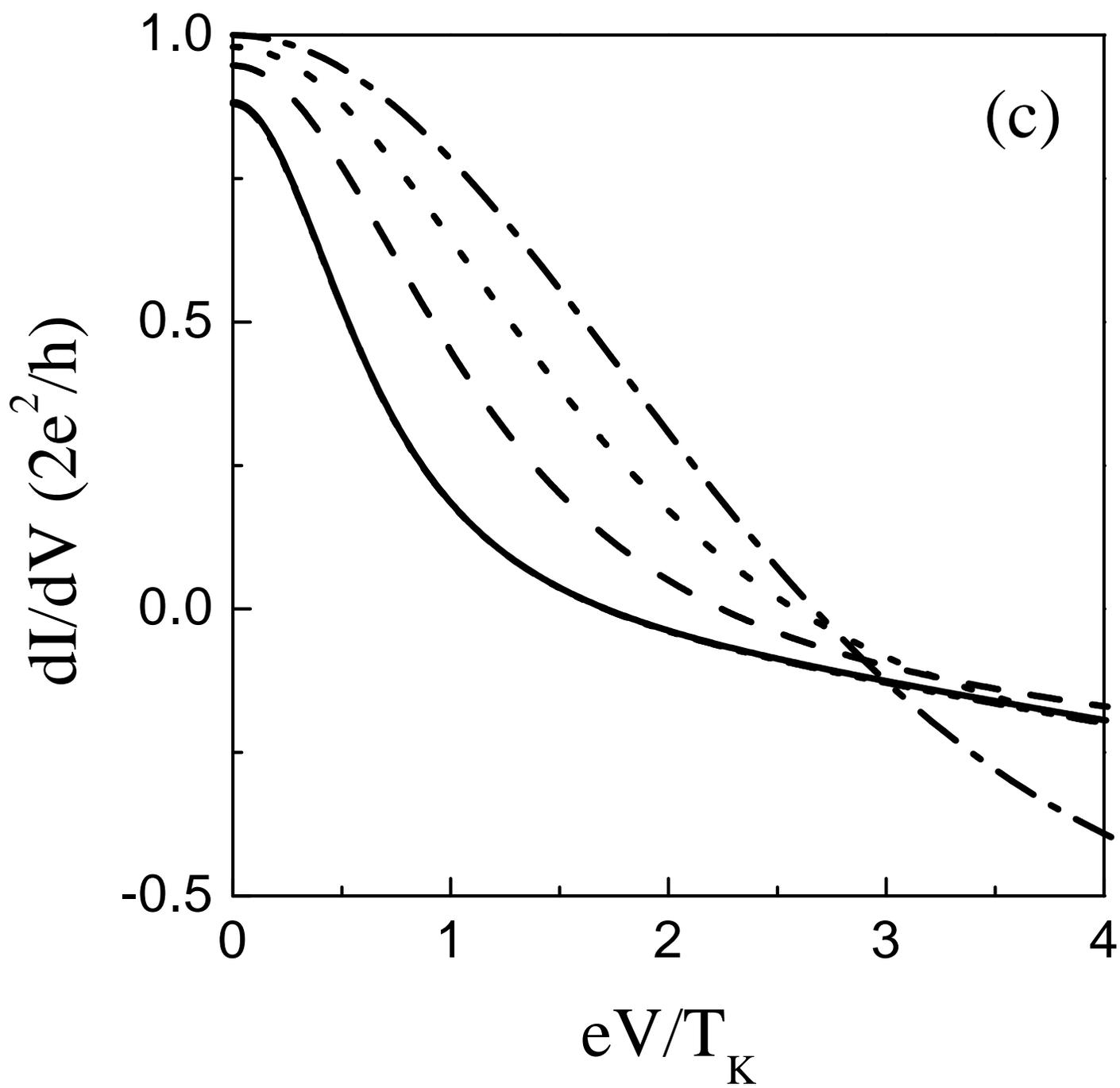
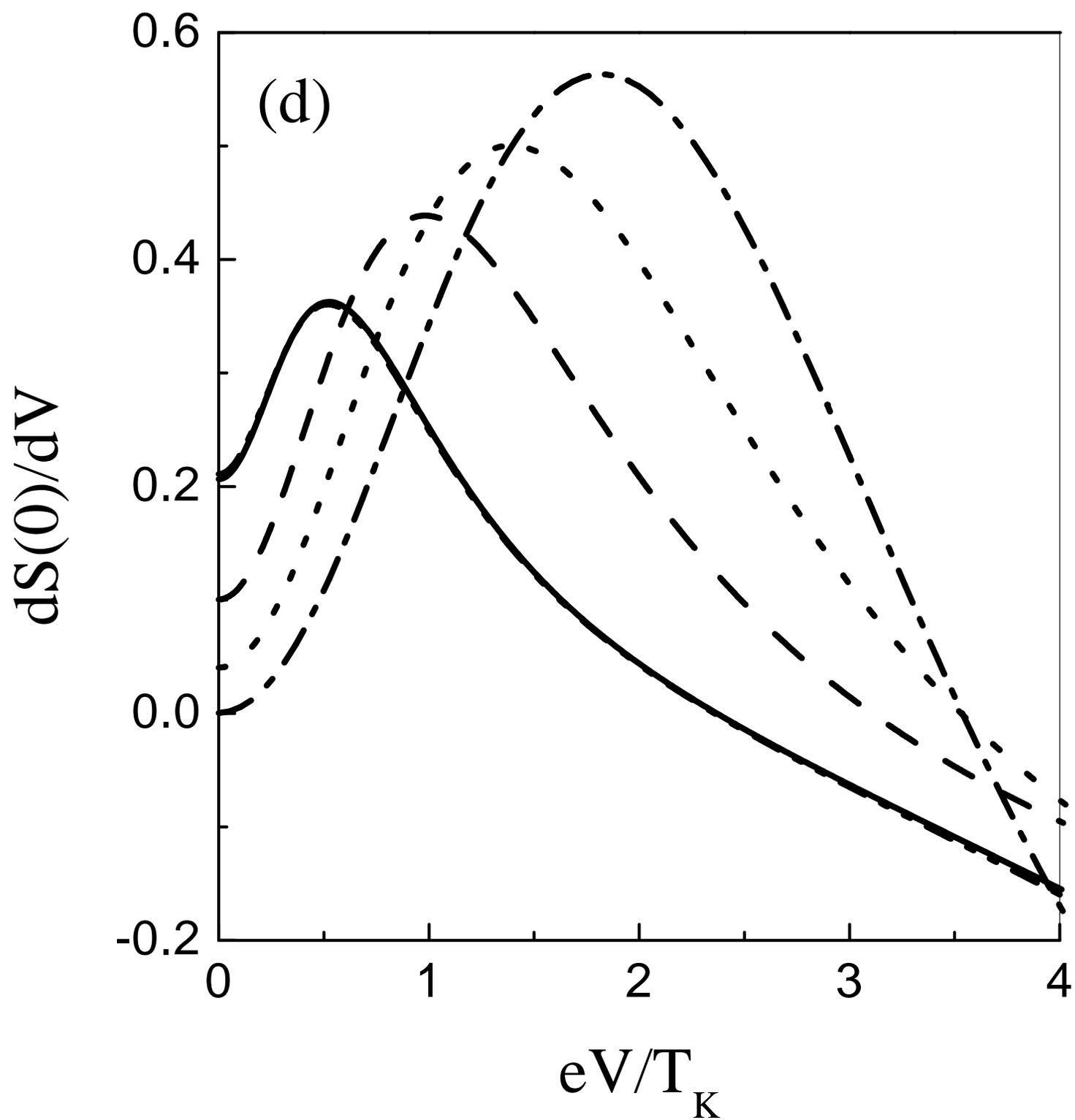
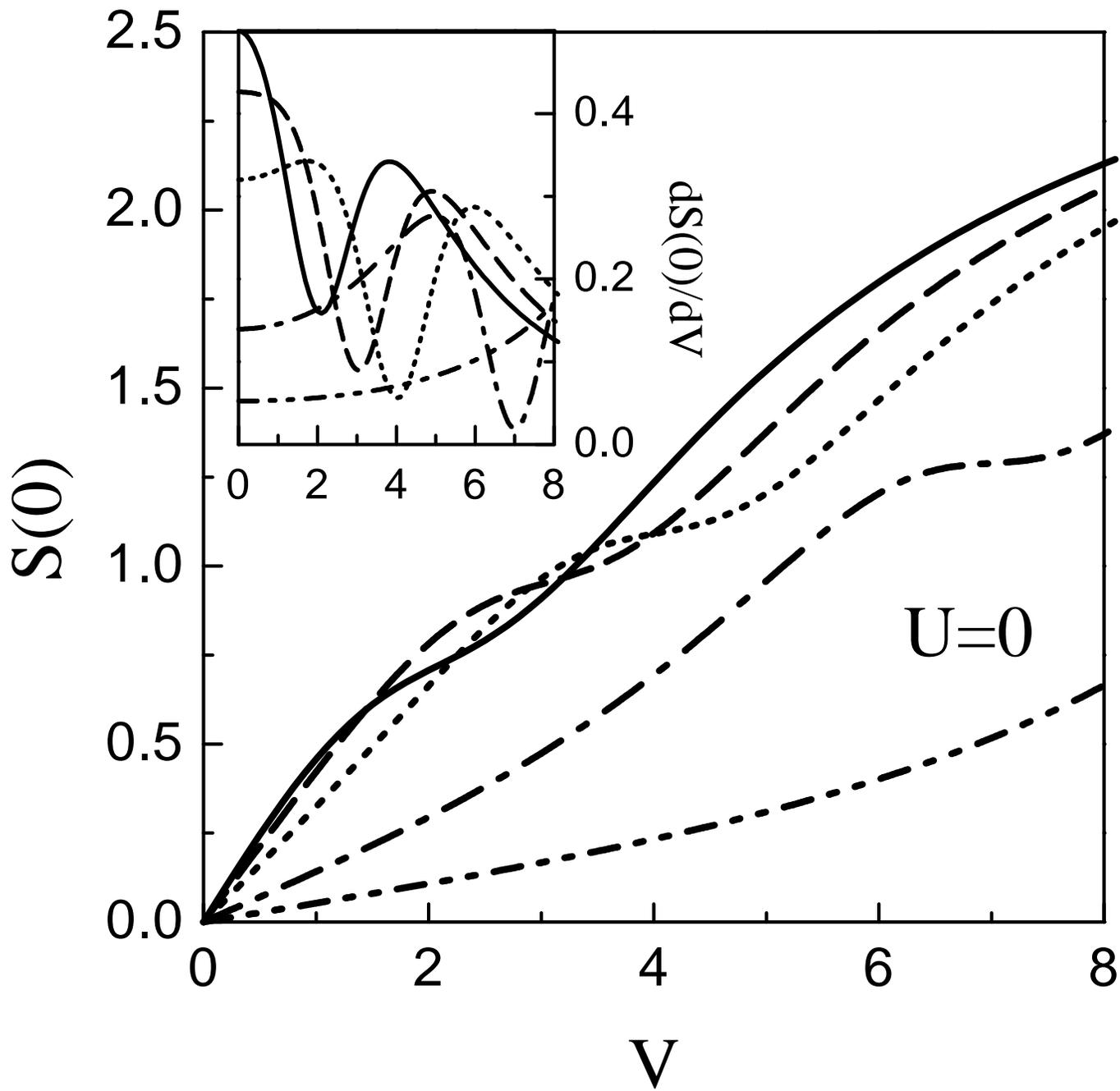
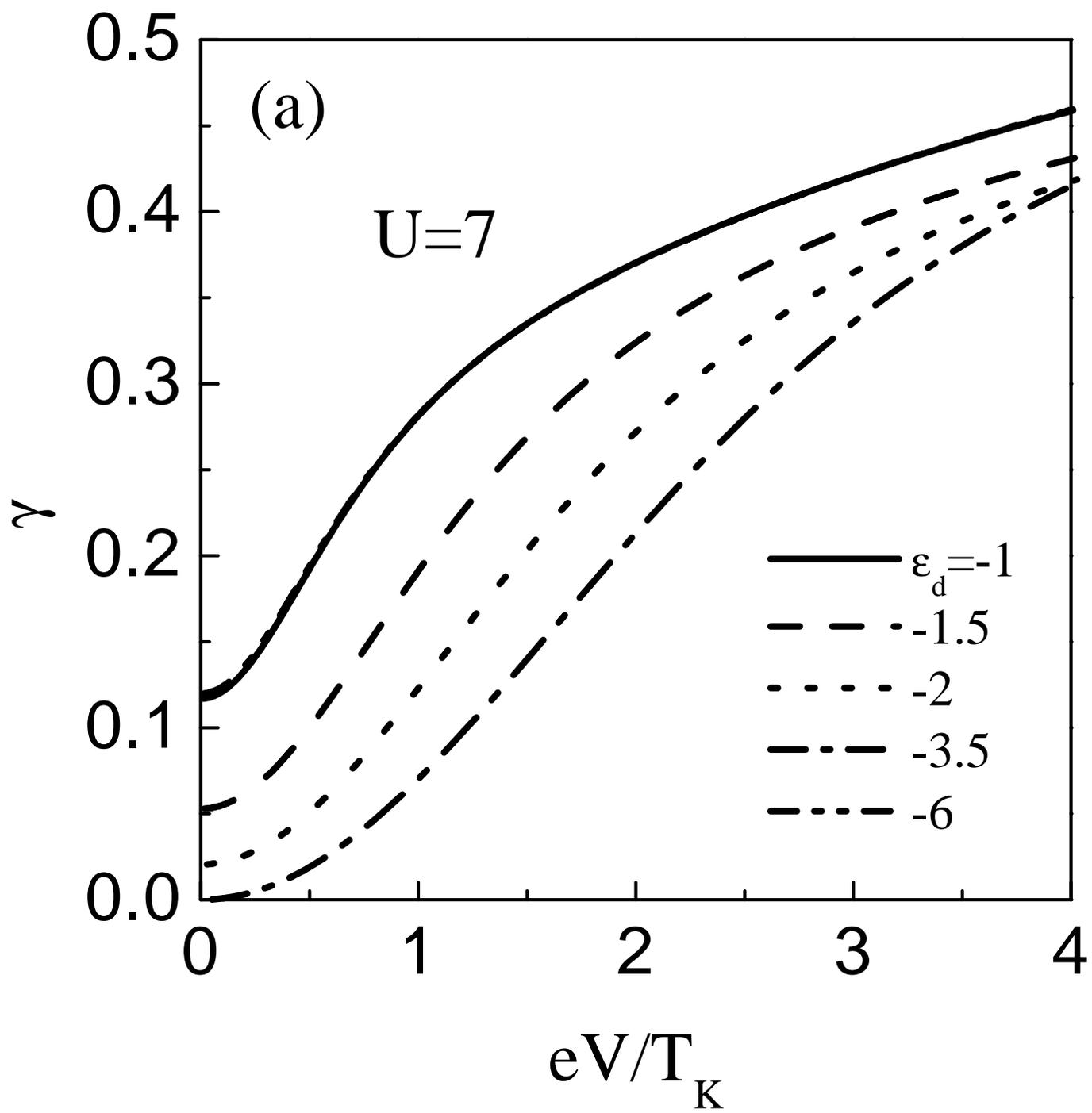
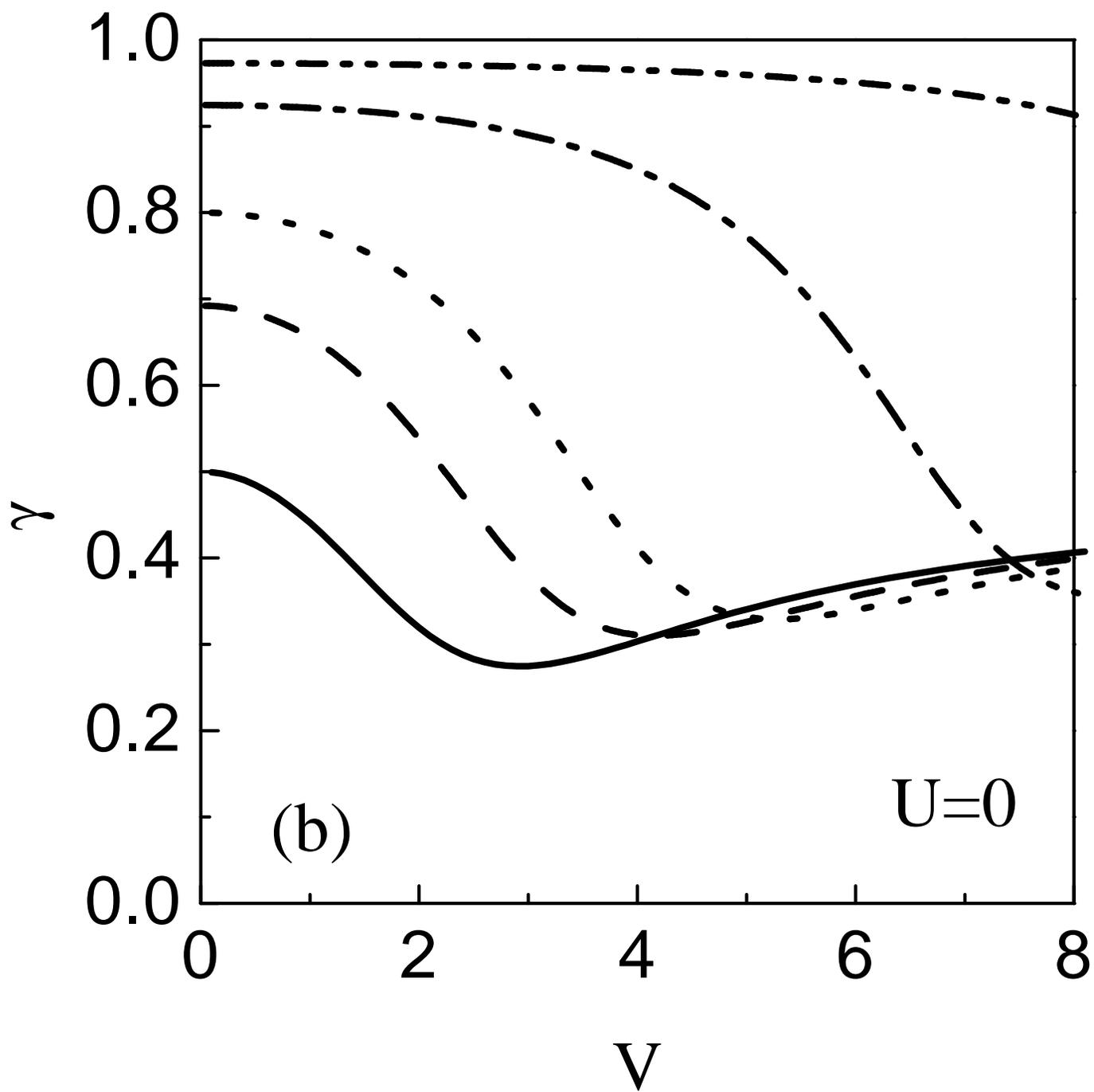
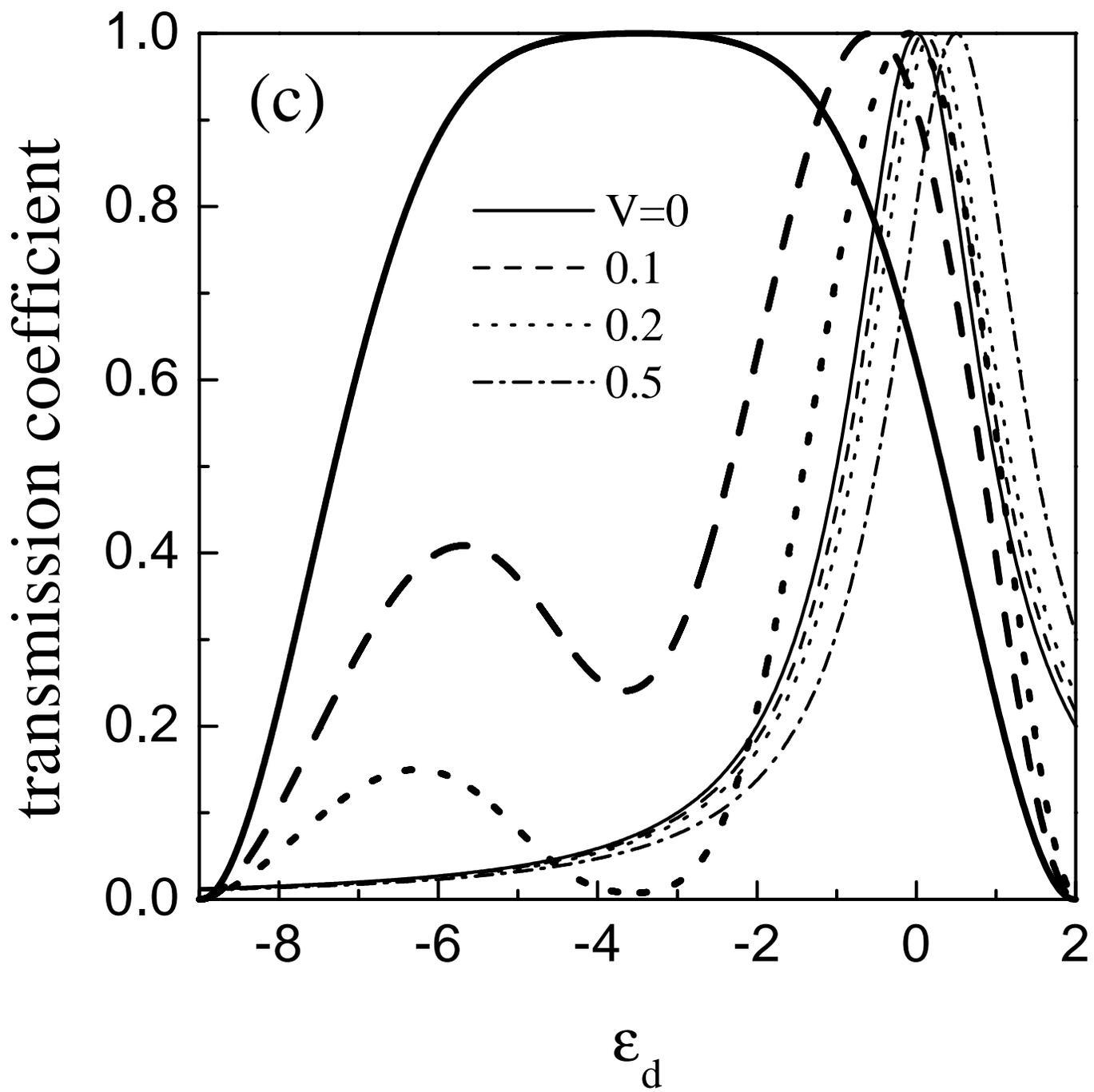
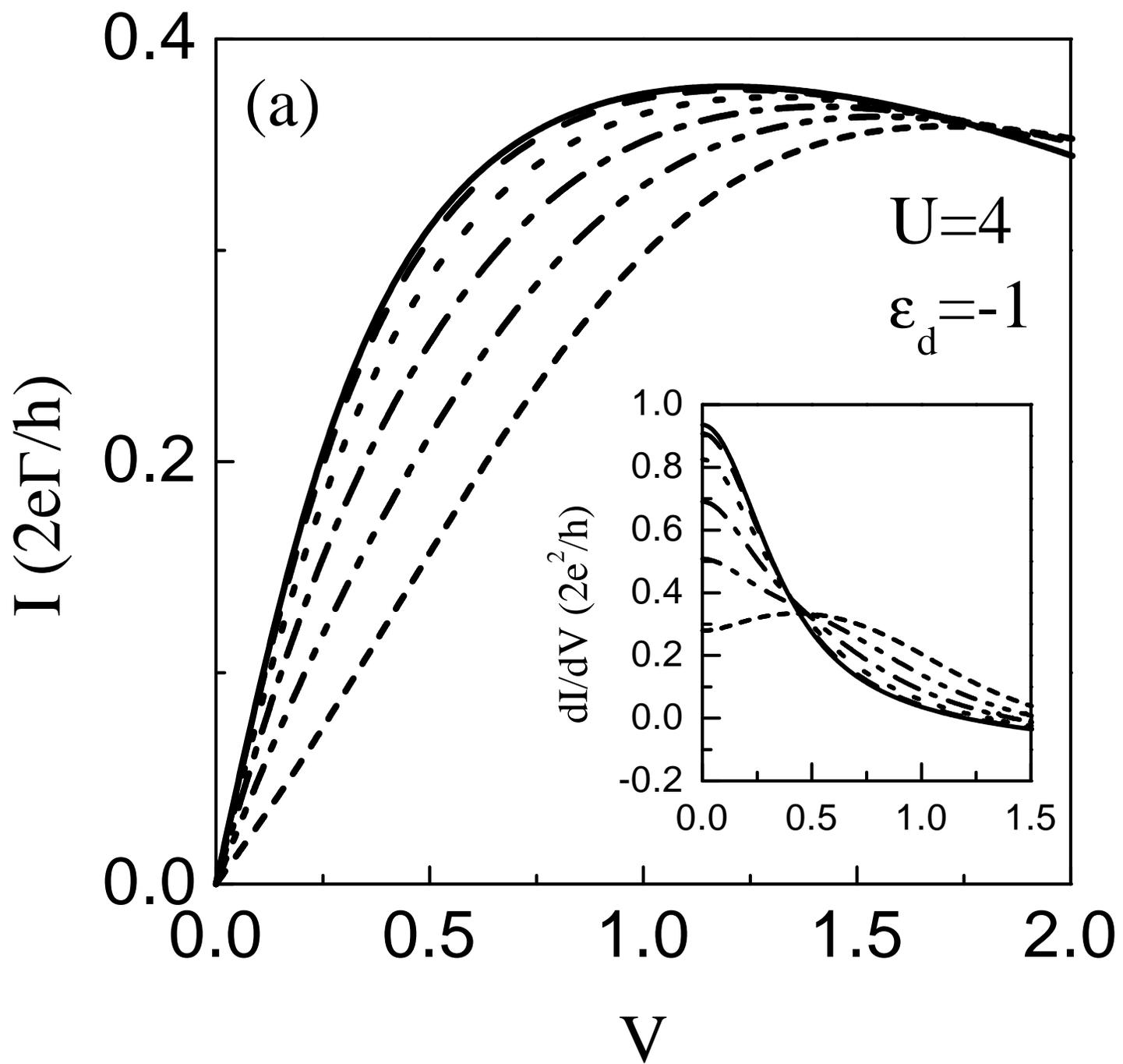
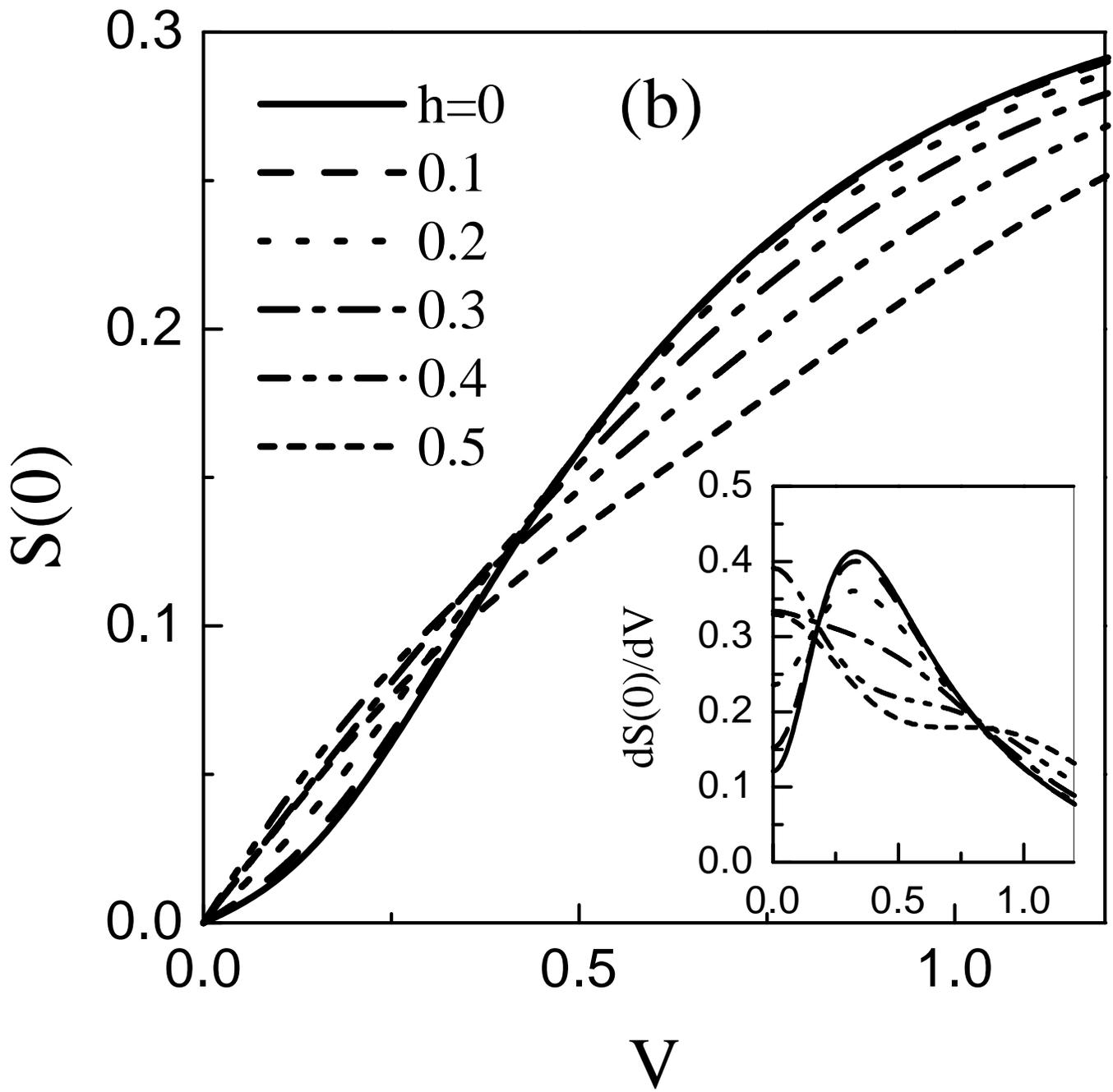
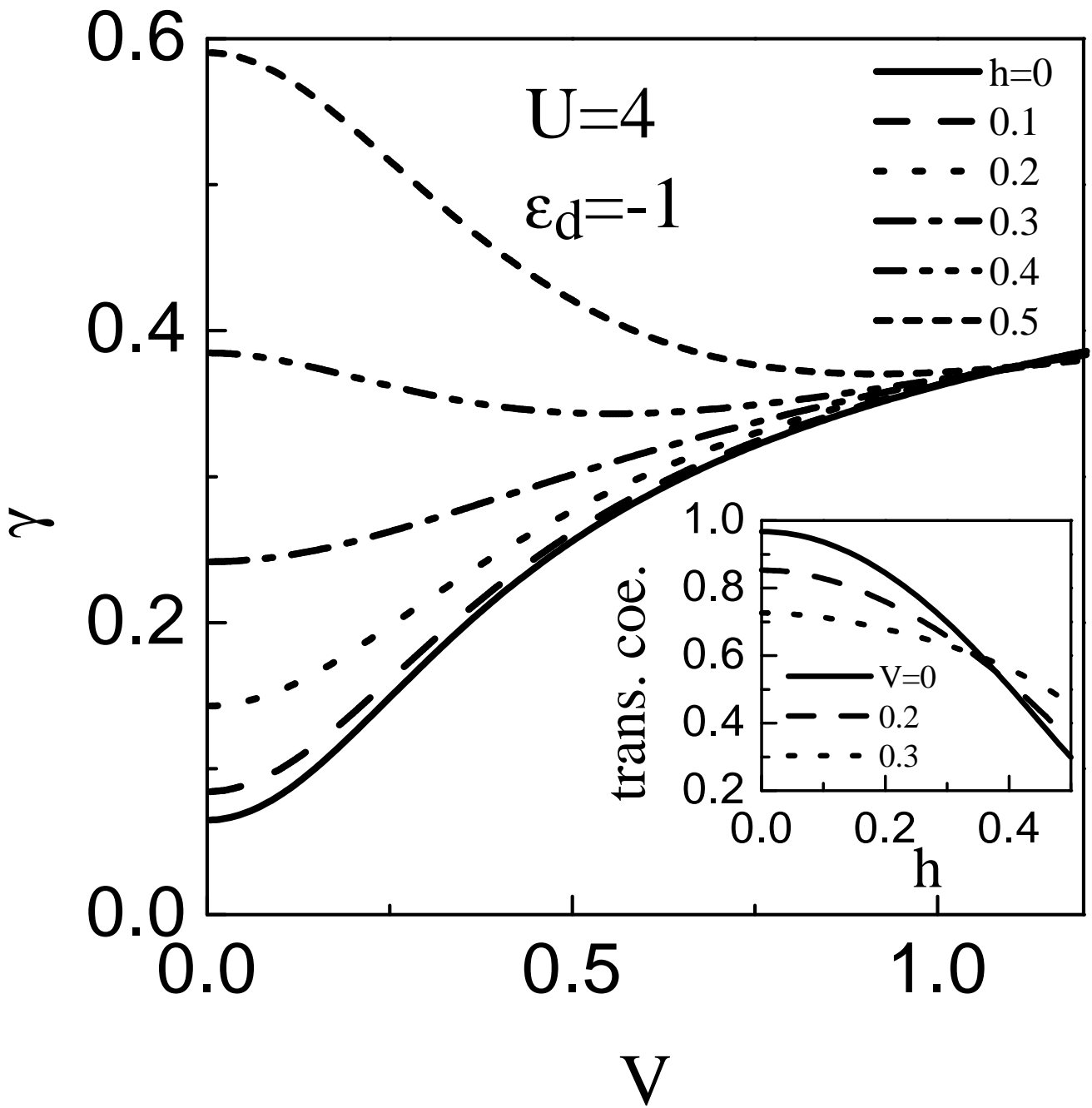
\end{document}